\documentclass[11pt]{article}

\usepackage[a4paper,margin=1in]{geometry}

\usepackage{amsmath,amssymb}

\usepackage{graphicx}

\usepackage{booktabs}

\usepackage[round,authoryear]{natbib}

\usepackage{listings}
\usepackage{xcolor}

\usepackage[hidelinks]{hyperref}

\definecolor{codegray}{rgb}{0.5,0.5,0.5}
\definecolor{backcolour}{rgb}{0.96,0.96,0.96}

\lstdefinestyle{Rstyle}{
    language=R,
    backgroundcolor=\color{backcolour},
    commentstyle=\itshape\color{codegray},
    keywordstyle=\bfseries\color{black},
    stringstyle=\color{black},
    numberstyle=\tiny\color{codegray},
    basicstyle=\ttfamily\footnotesize,
    breakatwhitespace=false,
    breaklines=true,
    captionpos=b,
    keepspaces=true,
    numbers=left,
    numbersep=5pt,
    showspaces=false,
    showstringspaces=false,
    showtabs=false,
    tabsize=2,
    frame=single
}

\title{%
Should the pharmaceutical industry consider using high-order
autoregressive models when modelling panel data?
}

\author{%
Dan Jackson
\and
Fanni Zhang
\and
Abdul-Azeez Ganiyu
\and
Rose Baker
}

\date{}

\begin{document}

\maketitle

\begin{abstract}
Longitudinal data, often in the form of panel data, are commonly encountered in the pharmaceutical industry. However it is not always possible to fit an unstructured covariance matrix when using the Mixed Model for Repeated Measures (MMRM). Here we argue that high-order autoregressive models provide a useful framework when more general correlation structures, such as the unstructured and/or Toeplitz, are unfeasible. The autoregressive model of order one is often suggested but, in our experience, higher orders are seldom considered. We present the case for using autoregressive models, with potentially high orders, as a flexible framework to model correlations in panel data, when more general structures are hard to identify.
\end{abstract}

\noindent\textbf{Keywords:}
Covariance matrix; Longitudinal data; Mixed Model for Repeated Measures;
Yule--Walker equations

\section{Introduction}

Longitudinal data, in the form of  panel data, are frequently encountered where the same outcome data are collected, at regular intervals, 
from a relatively large 
number of patients. One common application is where patient reported outcome (PRO) data are collected at a series of visits and 
numerous other applications may be encountered \citep{lafayette_phase_2026, yan_efficacy_2024, ganmaa_vitamin_2024}. 
A standard method for analysing panel 
data is the Mixed Model for Repeated Measures \citep{laird_random-effects_1982, garcia-hernandez_mmrm_2020, andersen_practical_2013, qian_comprehensive_2021, siddiqui_mmrm_2009} (MMRM). 
Even more sophisticated, and related, approaches are also available, 
for example joint models \citep{daza_joint_2024}, non-linear models \citep{yuan_novel_2020, wang_novel_2024} and models 
for ordinal data  \citep{demirtas_mixed_nodate, krol_location-scale_2026} but in our experience their use is much less widespread than the MMRM. 

A suitable covariance structure must be adopted when using the MMRM. Fully unstructured covariance matrices allow this structure 
to be as flexible as possible but are hard to identify when the number of time points is large. 
Simpler correlation structures are then 
typically used, where the autoregressive model of order one is a popular choice. However our position is that higher orders, that are seldom considered in practice, 
provide a useful framework when more general models are unfeasible. We aim to present a convincing case to support this position, and so to 
encourage others to consider using this type of model.

\section{The MMRM}
Let $\mathbf{y}_i$ be the column vector of 
responses from the $i$th subject. Let  $\mathbf{\beta}$ be the column vector of regression parameters and $\mathbf{X}_i$ be the design 
matrix linking $\mathbf{\beta}$ to $\mathbf{y}_i$. In a clinical trial 
setting we often wish to directly model the covariance structure of the responses and so do not include any random-effects. The motivation for this 
simplification may appear puzzling to the uninitiated, but it has the merit of avoiding unnecessary complications. For example, 
\cite{garcia-hernandez_mmrm_2020} remark that   `Curiously, the term MMRM is mostly used for a likelihood-based multivariate normal linear model without any random effects'.   \cite{Fairclough2010QoLTrials2e} remarks that `While this is not strictly a mixed-effects model [the MMRM without random-effects] 
has strongly been associated with that term'.  \cite{Pinheiro2000MixedEffects} refer to this as an extended linear model with no random-effects.  In the absence of random-effects,  the model becomes 
\begin{equation}
\label{model}
\mathbf{y}_i = \mathbf{X}_i \mathbf{\beta} + \mathbf{\epsilon}_i
\end{equation} 
We follow \cite{Pinheiro2000MixedEffects} in describing how the correlation structure is implemented in practice. On their pages 204-205 they define 
the statistical error in model (\ref{model}) to be  
$  \mathbf{\epsilon}_i \sim N(\mathbf{0}, \sigma^2 \mathbf{\Lambda}_i )$ where   $\mathbf{\Lambda}_i = \mathbf{V}_i \mathbf{C}_i \mathbf{V}_i$, $\mathbf{V}_i$ is diagonal and $\mathbf{C}_i$ 
is a correlation matrix.  The variance and correlation structures 
of the statistical errors  $\mathbf{\epsilon}_i$ in model (\ref{model}) are therefore described by $\mathbf{V}_i$ and $\mathbf{C}_i$, respectively. As explained by  \cite{Pinheiro2000MixedEffects}, $\mathbf{V}_i$ is not uniquely defined and so to ensure uniqueness we require all its diagonal entries to be positive. 
We usually assume the same covariance structure for all subjects. Our main focus is the 
statistical modelling required for the correlation structure.

\section{Autoregressive models}
We now examine autoregressive models in detail. 
We continue to follow \cite{Pinheiro2000MixedEffects} by, for simplicity, using the notation $\epsilon_t$ 
to refer to an observation at time $t$. Then an autoregressive model of order  $w$ assumes that
\begin{equation}
\label{AR}
 \epsilon_t=\phi_1  \epsilon_{t-1} + \phi_2  \epsilon_{t-2} +  \cdots + \phi_w  \epsilon_{t-w} +a_t
\end{equation}
where $a_t$ is an independent homoscedastic `white noise' term, centred at zero. Conceptually, the observation $\epsilon_t$ is assumed to depend 
on the $w$ observations that immediately preceded it, and a statistical error $a_t$. We further assume that model (\ref{AR}) is stationary, with mean of zero. 
A full discussion of standard methods for the analysis of time series is beyond the scope of this paper, but briefly 
a stationary time series is one whose statistical behaviour does not change over time. 

\subsection{The Yule-Walker equations}
The long-established Yule-Walker equations relate the  coefficients of a stationary  autoregressive model to the correlations. 
Writing $\rho_p$ for the correlation between an observation and the one $p$ time-periods earlier (the autocorrelation at lag $p$), these equations take 
matrix form
\begin{equation}
\label{total}
\left( 
 \begin{matrix}
    \rho_1 \\
    \rho_2 \\ 
    \vdots \\
    \rho_{w-1} \\ 
    \rho_w \\
  \end{matrix} \right)  = 
  \left( 
 \begin{matrix}
   \rho_0 & \rho_1 & \rho_2 & \cdots & \rho_{w-2} & \rho_{w-1} \\
   \rho_1 & \rho_0 & \rho_1 & \cdots & \rho_{w-3} & \rho_{w-2}\\
   \vdots & \vdots  & \vdots & \vdots & \vdots  & \vdots \\ 
   \rho_{w-2} & \rho_{w-3} & \rho_{w-4} & \cdots & \rho_0 & \rho_1 \\
   \rho_{w-1} & \rho_{w-2} & \rho_{w-3} & \cdots & \rho_1 & \rho_0 \\
  \end{matrix} \right)
  \left( 
 \begin{matrix}
    \phi_1 \\
    \phi_2 \\ 
    \vdots \\
    \phi_{w-1} \\ 
    \phi_w \\
  \end{matrix} \right)
\end{equation}
where $\rho_0=1$, so that all entries along the main diagonal of the matrix in (\ref{total}) are one.

\subsection{Using high order autoregressive models as a framework for modelling correlations for panel data}

 A key observation is that, because all correlations 
depend only on the lag (i.e. the number of time periods between observations), the matrix in (\ref{total})  
is a perfectly general Toeplitz correlation structure across the $w$ visits. Suppose, in the context of panel data, that there are $v$ visits, so that $w=v-1$ is the number of 
correlations needed 
to specify the Toeplitz correlation structure. Beginning with this 
Toeplitz structure, from equation  (\ref{total}) we can 
derive the coefficients of a corresponding stationary autoregressive model, of order $w$, that results in exactly the same set of correlations over the $v$ visits.  
Equation (\ref{total}) 
will always be uniquely solvable 
for any valid Toeplitz correlation structure  (except for in extreme and artificial situations, for example where some correlations are one). 

Autoregressive models therefore provide a useful conceptual framework when the Toeplitz correlation structure is not possible to adequately identify. 
This is because we can begin with a stationary AR($w$), corresponding to any given general Toeplitz correlation structure, 
and gradually reduce the order 
of the autoregressive model until it is adequately identified. We can therefore use autoregressive models as a framework to approximate the Toeplitz  structure, to the greatest 
extent feasible.

\section{Simulation study}
A simple scenario was used to illustrate the benefit of using high order autoregressive models. Here data were 
repeatedly simulated for 200 patients (100 in each of two treatment groups, treatment versus control) 
over 10 visits. 
The mean outcome for control group patients was zero for all 10 visits and for treatment group these means were all one. 
There is therefore a constant treatment effect of one for all visits.

Data were simulated using a Toeplitz correlation structure, where variances of all outcome data was one, with correlations 
$\rho_1=0.7, \rho_2=0.6, \rho_3=0.5, \rho_4=0.45, \rho_5=0.42, \rho_6=0.4, \rho_7=0.38, \rho_8=0.36, \rho_9=0.35$.
There is therefore a quick, exponential-like, decay in correlations at lags one, two and three. However this decay then slows down markedly 
and relatively strong correlations at very long lags are present. An AR(1) model will not be able to 
describe these data well.

Data at visit 1 were complete, but smaller 
outcomes at this first visit were assumed to be associated with a greater probability of patient drop out. In every simulated dataset, and within each arm, the outcome data at visit 1 
were sorted and those with the smallest value were taken to drop out first. Monotone drop out (where, once a subject has 
dropped out, they cease to provide outcome data at later visits) 
was induced in each arm. 10\% of treatment group patients, and 50\% 
of control group patients, dropped out at visit 2, respectively. These percentages were increased 
to 20\% and 60\% at visit 4, 30\% and 70\% at visit 6, and 50\% and 90\% at visit 8. Drop out rates (common 
across both arms) of 
95\% and 98\% were invoked at visits 9 and 10, respectively. 

This is a missing at random (MAR) scenario. Different subject specific missing data indicators are generated in 
every simulated dataset, where these indicators depend only 
on observed data (at visit 1). Hence likelihood based analyses, using an adequate 
statistical model, can be expected to be unbiased. The limitations of the AR(1) model, combined with the differential 
drop out rates across treatment groups, was expected to result in biases that the more sophisticated AR(5) model could reduce. 

A simple model was used for the fixed effects in $\mathbf{X}_i$, where different intercepts and treatment effects 
were allowed at each time point but no other variables were adjusted for. Four different correlation structures were explored: unstructured, Toeplitz, 
and autoregressive models of orders 5 and 1.  1000 simulated datasets were produced.

\subsection{Results}
The estimation of the unstructured 
covariance matrix failed in every instance. The Toeplitz
correlation structure failed in 77\% of simulated datasets. Hence simpler correlation structures are necessary.

The two 
autoregressive models encountered no estimation failures. In Figure 1 we show the average estimated treatment effect for all 10 visits, under both the  AR(1) and AR(5) models. 95\% 
Monte Carlo uncertainty intervals are also shown, where these intervals are completely 
	hidden by the majority of plotting points, indicating that the corresponding Monte Carlo error 
	is negligible.  There is no evidence of an important bias at visits one and two 
under the AR(1) model, which can be explained because data at visit 1 are complete and this model can accommodate the early 
exponential-like correlation decay at visit 2. However the AR(1) model results in notable (around 0.1 to 0.25) negative 
biases in visits 3 to 10.  The AR(5) model largely
eliminated these negative biases, despite also being incorrectly specified.

\begin{figure}
    \centering
    \includegraphics[width=0.8\textwidth]{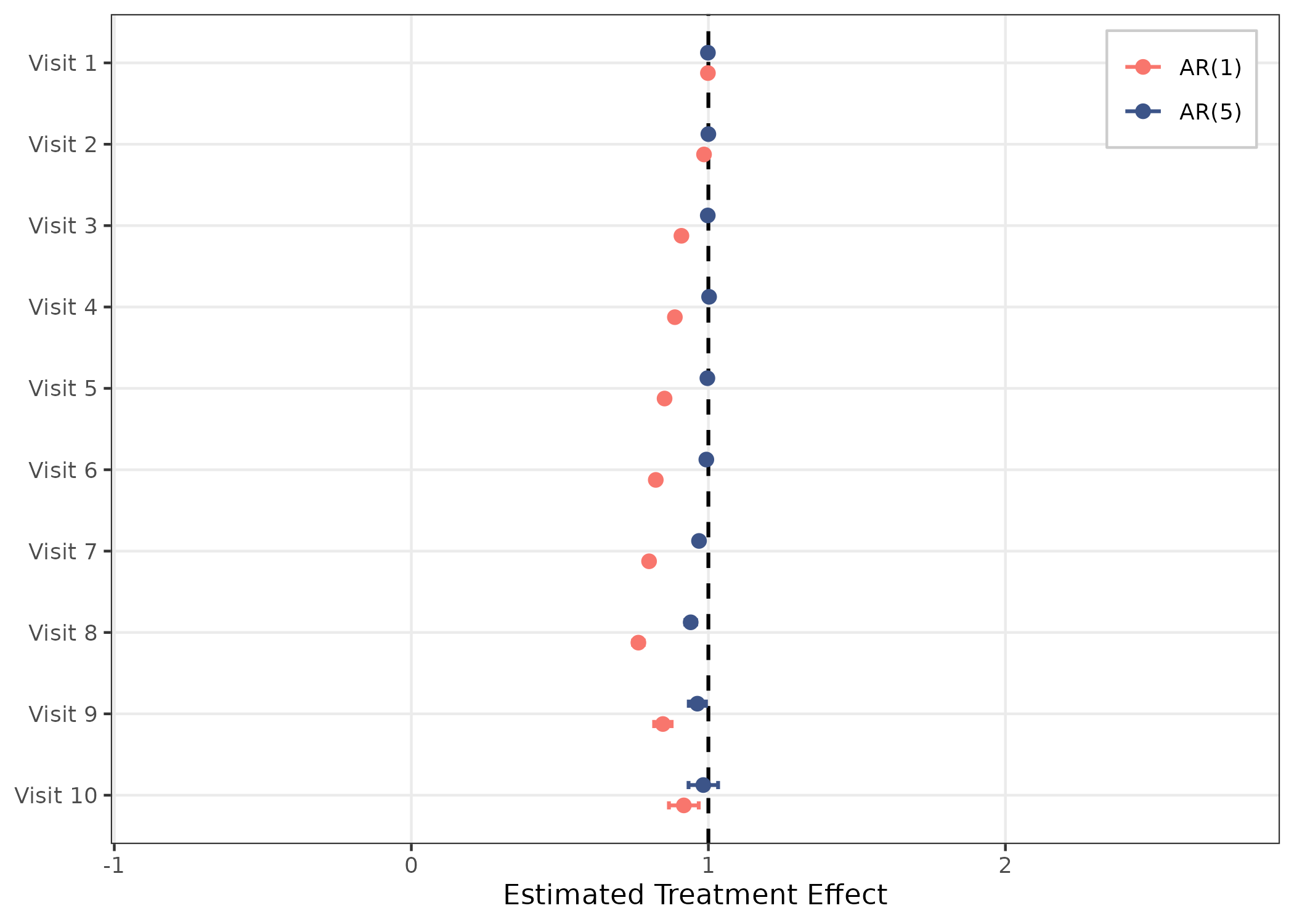}
    \caption{Results for the simulation study. The average treatment effects for all 10 visits are shown, under the AR(1) and AR(5) models, across all 1000 simulated datasets. 95\% Monte Carlo intervals are also shown, where these intervals are completely hidden by the majority of plotting points, indicating that the corresponding Monte Carlo error is negligible.}
    \label{fig:2}
\end{figure}

\section{Discussion}
This work was partly motivated by some of the authors encountering statistical analysis plans (SAPs), where a sequential correlation modelling strategy was proposed. In these plans, the  
unstructured correlation 
matrix was to be attempted first. If this proved unfeasible, then a Toeplitz would be used instead. Next, 
an autoregressive model of order one would be used. In our experience, this type of strategy may lead to overly, and 
unnecessarily, simple models being fitted.  We seek to improve 
current practice by advocating the use of higher order autoregressive models. 

We hope that we have made a convincing case that  high order autoregressive models provide a flexible framework for 
modelling correlations in sparse panel data. This framework allows us to model the associations between outcome data, from the same 
subject, with as much sophistication as the data will allow.

\section*{Acknowledgements}

We would like to acknowledge Dan James for useful discussions. 

\bibliographystyle{abbrvnat}
\bibliography{references2}

@article{qian_comprehensive_2021,
	title = {Comprehensive review of statistical methods for analysing patient-reported outcomes ({PROs}) used as primary outcomes in randomised controlled trials ({RCTs}) published by the {UK}’s {Health} {Technology} {Assessment} ({HTA}) journal (1997–2020)},
	copyright = {© Author(s) (or their employer(s)) 2021. Re-use permitted under CC BY. Published by BMJ.. https://creativecommons.org/licenses/by/4.0/This is an open access article distributed in accordance with the Creative Commons Attribution 4.0 Unported (CC BY 4.0) license, which permits others to copy, redistribute, remix, transform and build upon this work for any purpose, provided the original work is properly cited, a link to the licence is given, and indication of whether changes were made. See: https://creativecommons.org/licenses/by/4.0/.},
	url = {},
	doi = {},
	volume = {11},
	language = {en},
	urldate = {2026-03-13},
	journal = {BMJ Open},
	author = {Qian, Yirui and Walters, Stephen J. and Jacques, Richard and Flight, Laura},
	month = sep,
	year = {2021},
	note = {},
	pages = {e051673},
}

@article{daza_joint_2024,
	title = {Joint models inform the longitudinal assessment of patient-reported outcomes in clinical trials: a simulation study and secondary analysis of the restrictive {Vs}. liberal fluid therapy for major abdominal surgery ({RELIEF}) randomized controlled trial},
	volume = {176},
	issn = {1878-5921},
	shorttitle = {Joint models inform the longitudinal assessment of patient-reported outcomes in clinical trials},
	doi = {},
	language = {eng},
	journal = {Journal of Clinical Epidemiology},
	author = {Daza, Julian F. and Mitani, Aya A. and Alibhai, Shabbir M. H. and Smith, Peter M. and Kennedy, Erin D. and Shulman, Mark A. and Myles, Paul S. and Wijeysundera, Duminda N.},
	month = dec,
	year = {2024},
	pmid = {39389273},
	pages = {111553},
}

@article{krol_location-scale_2026,
	title = {Location-{Scale} {Latent} {Process} {Model} for {Repeated} {Ordinal} {Patient}-{Reported} {Outcomes}},
	volume = {45},
	copyright = {© 2026 John Wiley \& Sons Ltd.},
	issn = {1097-0258},
	url = {},
	doi = {},
	language = {en},
	number = {},
	urldate = {2026-03-13},
	journal = {Statistics in Medicine},
	author = {Król, Agnieszka and Palmér, Robert and Leander, Jacob and Proust-Lima, Cécile and Jauhiainen, Alexandra},
	year = {2026},
	note = {},
	pages = {e70482},
}

@article{demirtas_mixed_nodate,
	title = {A mixed ordinal location scale model for analysis of ecological momentary assessment ({EMA}) data},
	volume = {2},
	issn = {1938-7997},
	url = {},
	doi = {},
	language = {EN},
	number = {4},
	urldate = {2026-03-13},
	journal = {Statistics and Its Interface},
	author = {Demirtas, Hakan and Hedeker, Donald and Mermelstein, Robin J.},
	note = {},
	pages = {391--401},
}

@article{andersen_practical_2013,
	title = {On the practical application of mixed effects models for repeated measures to clinical trial data},
	volume = {12},
	copyright = {Copyright © 2012 John Wiley \& Sons, Ltd.},
	issn = {1539-1612},
	url = {},
	doi = {},
	language = {en},
	number = {1},
	urldate = {2026-03-13},
	journal = {Pharmaceutical Statistics},
	author = {Andersen, Scott W. and Millen, Brian A.},
	year = {2013},
	note = {},
	pages = {7--16},
}

@article{garcia-hernandez_mmrm_2020,
	title = {{MMRM} vs joint modeling of longitudinal responses and time to study drug discontinuation in clinical trials using a “de jure” estimand},
	volume = {19},
	copyright = {© 2020 John Wiley \& Sons Ltd},
	issn = {1539-1612},
	url = {},
	doi = {},
	language = {en},
	number = {6},
	urldate = {2026-03-13},
	journal = {Pharmaceutical Statistics},
	author = {García-Hernandez, Alberto and Pérez, Teresa and Pardo, María del Carmen and Rizopoulos, Dimitris},
	year = {2020},
	note = {},
	pages = {909--927},
}

@article{siddiqui_mmrm_2009,
	title = {{MMRM} vs. {LOCF}: a comprehensive comparison based on simulation study and 25 {NDA} datasets},
	volume = {19},
	issn = {1520-5711},
	shorttitle = {{MMRM} vs. {LOCF}},
	doi = {},
	language = {eng},
	number = {2},
	journal = {Journal of Biopharmaceutical Statistics},
	author = {Siddiqui, Ohidul and Hung, H. M. James and O'Neill, Robert},
	year = {2009},
	pmid = {19212876},
	pages = {227--246},
}

@article{yuan_novel_2020,
	title = {A novel quantification of information for longitudinal data analyzed by mixed-effects modeling},
	volume = {19},
	copyright = {© 2020 John Wiley \& Sons Ltd},
	issn = {1539-1612},
	url = {},
	doi = {},
	language = {en},
	number = {4},
	urldate = {2026-03-13},
	journal = {Pharmaceutical Statistics},
	author = {Yuan, Min and Li, Yi and Yang, Yaning and Xu, Jinfeng and Tao, Fangbiao and Zhao, Liang and Zhou, Honghui and Pinheiro, Jose and Xu, Xu Steven},
	year = {2020},
	note = {},
	pages = {388--398},
}

@article{wang_novel_2024,
	title = {Novel non-linear models for clinical trial analysis with longitudinal data: {A} tutorial using {SAS} for both frequentist and {Bayesian} methods},
	volume = {43},
	copyright = {© 2024 John Wiley \& Sons Ltd.},
	issn = {1097-0258},
	shorttitle = {Novel non-linear models for clinical trial analysis with longitudinal data},
	url = {},
	doi = {},
	language = {en},
	number = {15},
	urldate = {2026-03-13},
	journal = {Statistics in Medicine},
	author = {Wang, Guoqiao and Wang, Whedy and Mangal, Brian and Liao, Yijie and Schneider, Lon and Li, Yan and Xiong, Chengjie and McDade, Eric and Kennedy, Richard and Bateman, Randall and Cutter, Gary},
	year = {2024},
	note = {},
	pages = {2987--3004},
}

@article{lafayette_phase_2026,
	title = {A {Phase} 3 {Trial} of {Atacicept} in {Patients} with {IgA} {Nephropathy}},
	volume = {394},
	issn = {0028-4793},
	url = {},
	doi = {},
	number = {7},
	urldate = {2026-03-13},
	journal = {New England Journal of Medicine},
	author = {Lafayette, Richard and Barbour, Sean J. and Brenner, Robert M. and Campbell, Kirk N. and Doan, Tom and Eren, Necmi and Floege, Jürgen and Jha, Vivekanand and Kim, Beom Seok and Liew, Adrian and Maes, Bart and Pal, Atanu and Pecoits-Filho, Roberto and Phoon, Richard K. S. and Rizk, Dana V. and Suzuki, Hitoshi and Tesař, Vladimir and Trimarchi, Hernán and Wei, Xuelian and Zhang, Hong and Barratt, Jonathan},
	month = feb,
	year = {2026},
	note = {},
	pages = {647--657},
}

@article{laird_random-effects_1982,
	title = {Random-effects models for longitudinal data},
	volume = {38},
	issn = {0006-341X},
	language = {eng},
	number = {4},
	journal = {Biometrics},
	author = {Laird, N. M. and Ware, J. H.},
	month = dec,
	year = {1982},
	pmid = {7168798},
	pages = {963--974},
}

@book{Pinheiro2000MixedEffects, author = {Pinheiro, Jos{'e} C. and Bates, Douglas M.}, title = {Mixed-Effects Models in S and S-PLUS}, publisher = {Springer}, address = {New York}, year = {2000}, series = {Statistics and Computing}, isbn = {0-387-98957-0} }

@book{Fairclough2010QoLTrials2e, author = {Fairclough, Diane L.}, title = {Design and Analysis of Quality of Life Studies in Clinical Trials}, publisher = {Chapman and Hall}, address = {London}, year = {2010}, series = {}, isbn = {} }

@article{yan_efficacy_2024,
	title = {Efficacy and safety of visepegenatide, a long-acting weekly {GLP}-1 receptor agonist as monotherapy in type 2 diabetes mellitus: a randomised, double-blind, parallel, placebo-controlled phase 3 trial},
	volume = {47},
	issn = {2666-6065},
	shorttitle = {Efficacy and safety of visepegenatide, a long-acting weekly {GLP}-1 receptor agonist as monotherapy in type 2 diabetes mellitus},
	url = {},
	doi = {},
	language = {English},
	urldate = {2026-04-08},
	journal = {},
	author = {Yan, Xiang and Ma, Jianhua and Liu, Yan and Wang, Xuhong and Li, Sheli and Yan, Shuang and Mo, Zhaohui and Zhu, Yikun and Lin, Jingna and Liu, Jie and Jia, Ying and Liu, Li and Ding, Ke and Xu, Michael and Zhou, Zhiguang},
	month = jun,
	year = {2024},
	note = {Publisher: Elsevier},
}

@article{ganmaa_vitamin_2024,
	title = {Vitamin {D} supplements for fracture prevention in schoolchildren in {Mongolia}: analysis of secondary outcomes from a multicentre, double-blind, randomised, placebo-controlled trial},
	volume = {12},
	issn = {2213-8587, 2213-8595},
	shorttitle = {Vitamin {D} supplements for fracture prevention in schoolchildren in {Mongolia}},
	url = {},
	doi = {},
	language = {English},
	number = {1},
	urldate = {2026-04-08},
	journal = {The Lancet Diabetes \& Endocrinology},
	author = {Ganmaa, Davaasambuu and Khudyakov, Polyna and Buyanjargal, Uyanga and Tserenkhuu, Enkhtsetseg and Erdenenbaatar, Sumiya and Achtai, Chuluun-Erdene and Yansanjav, Narankhuu and Delgererekh, Baigal and Ankhbat, Munkhzaya and Tsendjav, Enkhjargal and Ochirbat, Batbayar and Jargalsaikhan, Badamtsetseg and Enkhmaa, Davaasambuu and Martineau, Adrian R.},
	month = jan,
	year = {2024},
	pmid = {38048799},
	note = {Publisher: Elsevier},
	pages = {29--38},
}

\end{document}